\documentclass[aip,superscriptaddress,reprint]{revtex4-1}

\usepackage{graphicx}
\usepackage{dcolumn}
\usepackage{bm}
\usepackage{amsmath,amssymb}
\usepackage{units}

\begin{document}

\title{Independent dynamic acousto-mechanical and electrostatic control of individual quantum dots in a LiNbO$_{3}$-GaAs hybrid} 

\author{Jens Pustiowski}
\affiliation{Lehrstuhl f\"{u}r Experimentalphysik 1 and Augsburg Centre for Innovative Technologies (ACIT), Universit\"{a}t Augsburg, Universit\"{a}tstr. 1, 86159 Augsburg, Germany} 

\author{Kai M\"{u}ller}
\affiliation{Walter Schottky Institut and Physik Department, TU M\"{u}nchen, Am Coulombwall 4, 85748 Garching, Germany}
\affiliation{E. L. Ginzton Laboratory, Stanford University, Stanford, California 94305, United States}
\affiliation{Nanosystems Initiative Munich (NIM), Schellingstr. 4, 80799 M\"{u}nchen, Germany}

\author{Max Bichler}
\affiliation{Walter Schottky Institut and Physik Department, TU M\"{u}nchen, Am Coulombwall 4, 85748 Garching, Germany} 

\author{Gregor Koblm\"{u}ller}
\affiliation{Walter Schottky Institut and Physik Department, TU M\"{u}nchen, Am Coulombwall 4, 85748 Garching, Germany} 
\affiliation{Nanosystems Initiative Munich (NIM), Schellingstr. 4, 80799 M\"{u}nchen, Germany}

\author{Jonathan J. Finley }
\affiliation{Walter Schottky Institut and Physik Department, TU M\"{u}nchen, Am Coulombwall 4, 85748 Garching, Germany} 
\affiliation{Nanosystems Initiative Munich (NIM), Schellingstr. 4, 80799 M\"{u}nchen, Germany}

\author{Achim Wixforth}
\affiliation{Lehrstuhl f\"{u}r Experimentalphysik 1 and Augsburg Centre for Innovative Technologies (ACIT), Universit\"{a}t Augsburg, Universit\"{a}tstr. 1, 86159 Augsburg, Germany} 
\affiliation{Nanosystems Initiative Munich (NIM), Schellingstr. 4, 80799 M\"{u}nchen, Germany}

\author{Hubert J. Krenner}
\email{hubert.krenner@physik.uni-augsburg.de}
\affiliation{Lehrstuhl f\"{u}r Experimentalphysik 1 and Augsburg Centre for Innovative Technologies (ACIT), Universit\"{a}t Augsburg, Universit\"{a}tstr. 1, 86159 Augsburg, Germany} 
\affiliation{Nanosystems Initiative Munich (NIM), Schellingstr. 4, 80799 M\"{u}nchen, Germany}

\date{\today}

\begin{abstract}
We demonstrate tuning of single quantum dot emission lines by the combined action of the \emph{dynamic} acoustic field of a radio frequency surface acoustic wave and a \emph{static} electric field. Both tuning parameters are set all-electrically in a LiNbO$_{3}$-GaAs hybrid device. The surface acoustic wave is excited directly on the strong piezoelectric LiNbO$_{3}$ onto which a GaAs-based \emph{p-i-n} photodiode containing a single layer of quantum dots was epitaxially transferred. We demonstrate dynamic spectral tuning with bandwidths exceeding 3 meV of single quantum dot emission lines due to deformation potential coupling. The center energy of the dynamic spectral oscillation can be independently programmed simply by setting the bias voltage applied to the diode. 
\end{abstract}


\maketitle 
The random nature of the nucleation of epitaxial quantum dot (QD) nanosystems leads to an inhomogeneous broadening of the dots' optical properties. Thus, reversible post-growth tuning mechanisms of the QD emission energy and occupancy states have been developed over the past 20 years. Here, the most established tuning parameter is a static electric field which can be simply set by tuning the bias voltage applied to a diode structure with embedded QDs. This tuning mechanism is routinely employed to control the occupancy state \cite{Drexler:94,Warburton:00} and emission energy\cite{Fry:00a} of QDs or coherent quantum couplings in QD-molecules\cite{Krenner:05b}. More recently, static\cite{Seidl:06,Ding:10} and dynamic strain \cite{Gell:08,Bruggemann2011} fields have proven to efficiently and independently tune the confined electronic and excitonic states of a QD. However, for novel quantum-optoelectronic devices and quantum logic protocols a combination of static and dynamic tuning parameters is of paramount importance. This sparked the idea to combine both tuning mechanisms to achieve full control over the QD's optical properties\cite{Trotta2012} which was crucial for realizing a highly reliable source of polarization entangled photon pairs \cite{Trotta2012a}. Furthermore, to implement advanced quantum logic protocols based on Landau-Zener transitions in architectures based on optically active QDs\cite{Blattmann2014}, tuning has to be performed at radio frequencies (rf). In this frequency band surface acoustic waves (SAWs) are an ideal candidate for dynamic tuning since these are accompanied by dynamic strain and piezoelectric components. Thus, they allow for acousto-mechanical and acousto-electric control of semiconductor nanostructures \cite*{Rocke:97,Rotter:99a,Kinzel:11}, including in particular, the dynamic control of both the occupancy state \cite{Couto:09,Schulein2013,Weiss2014b} and emission energy \cite{Gell:08,Weiss2014a} of QDs. Because SAWs propagate almost dissipation-free over chip-scale distances, they natively address individual located along their propagation direction in parallel. \\Here we demonstrate independent control of single QD emitters by \emph{dynamic strain} and \emph{static electric field} tuning in a hybrid $\rm LiNbO_3-GaAs$ device fabricated by epitaxial lift-off and transfer\cite{Rotter:97b,Fuhrmann:10a}. Strain tuning is achieved by electrically exciting a SAW on a $\rm LiNbO_3$ host substrate which interacts with QDs embedded in the active layer of a $\rm GaAs$-based single QD-photodiode. We show that the latter allows for a global static electrical control of both the QD emission energy and occupancy state and the SAW provides a fast modulation around this statically defined center energy. \\

\begin{figure}[htb]
\vspace{0.3cm}
\includegraphics[width=.90\linewidth]{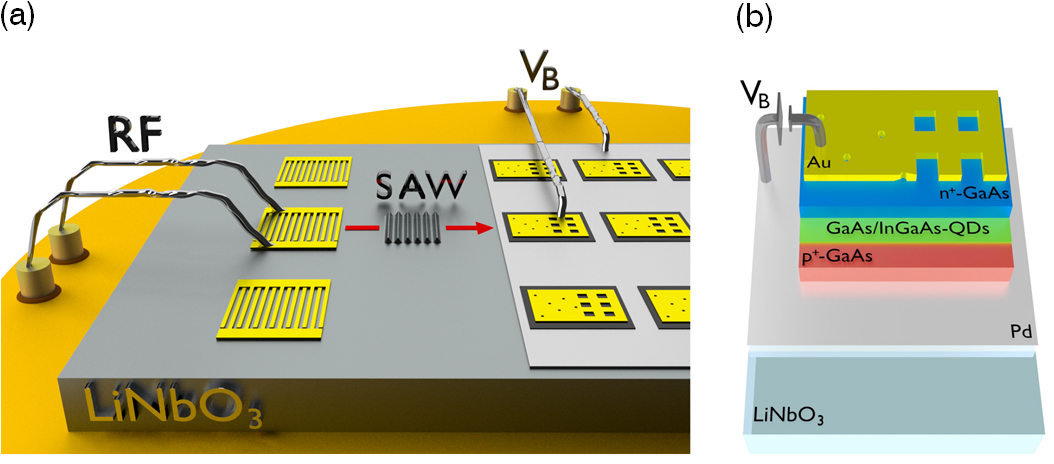}%
\caption{(Color online) (a) Schematic of our hybrid device consisting of a LiNbO$_{3}$ SAW-chip equipped with IDTs and a single QD \emph{p-i-n}-photodiode. (b) Layer sequence of the optically active epitaxially transferred semiconductor film bonded to the LiNbO$_{3}$ substrate via a Pd adhesion layer.}\label{FIG1}
\end{figure}

Our hybrid device is shown schematically in Fig.\ref{FIG1}(a). It consists of a LiNbO$_3$ host substrate and an optically active GaAs based quantum dot structure as illustrated in Fig.\ref{FIG1}(b). 
This structure was grown by molecular beam epitaxy (MBE) on a semi-insulating GaAs (100) substrate. After growth of a GaAs buffer layer, we deposited a \unit[100]{nm} thick AlAs sacrificial layer for a selective wet chemical etching step. On top of this sacrificial layer, we grew \unit[200]{nm} heavily p-doped GaAs followed by an undoped \unit[35]{nm} GaAs buffer. Self-assembled QDs were formed by depositing 5 ML of In$_{0.5}$Ga$_{0.5}$As, which were subsequentially overgrown by \unit[280]{nm} intrinsic GaAs and a \unit[200]{nm} heavily n-doped GaAs contact. To generate SAWs on the chip, Ti/Au interdigital transducers (IDTs) with a resonant frequency $f_{SAW}=\unit[292]{MHz}$ (acoustic wavelength $\lambda_{SAW}=\unit[13.6]{\mu m}$) were fabricated prior transfer of the semiconductor film on a \unit[128]{$^\circ$} rot YX LiNbO$_3$ substrate. A \unit[50]{nm} palladium (Pd) metallization was deposited at the later position of the semiconductor film. After fabrication of a \unit[200]{nm} Au n-side contact equipped with shadow mask with \unit[$\approx$ 1]{$\mu$m} diameter apertures to isolate single QDs. We epitaxially lifted off the diode structure from the GaAs substrate by selective HF-etching of the sacrificial layer\cite{Yablonovitch1987,Yablonovitch1990}. This \unit[715]{nm} thick film was transferred onto the SAW-chip with the p-doped side on forming both a strong mechanical bond and good electrical contact\cite{Yablonovitch1991} to the Pd layer. After transfer, we electrically isolated individual photodiodes by etching mesas.  
\\

Our experiments were performed in a liquid helium flow cold-finger cryostat at low temperature ($T=\unit[10]{K}$) using a conventional micro-photoluminescence ($\mu$-PL) setup. For quasi-resonant photogeneration of charge carriers we used an externally triggered pulsed diode laser emitting $\tau_{laser}\approx\unit[90]{ps}$ pulses of a wavelength $\lambda_{laser}= \unit[850]{nm}$. The laser was focused by a 50$\times$ microscope objective to one of the aforementioned microapertures and the emitted PL of the QD was collected by the same objective and dispersed by a \unit[0.5]{m} grating monochromator. Time-integrated detection was performed using a liquid N$_2$ cooled Si-charge coupled device (CCD). The SAW was excited in pulsed mode ($f_{rep}=\unit[100]{kHz}$, on/off duty cycle 1:9,) to reduce spurious heating of the sample. To record the time-averaged SAW-modulation of the QD emission, we set $f_{SAW}\neq n\cdot f_{laser}$, with $f_{laser}$ being the laser repetition rate. In contrast, for time domain studies, we employed stroboscopic excitation $(f_{SAW}=n\cdot f_{laser})$ and recorded time-integrated spectra for a fixed temporal delay $\tau_{delay}$ (relative phase $\varphi$) between laser and SAW over two full cycles from $-T_{SAW}\leq \tau_{delay} < +T_{SAW}$ ($-2\pi \leq \varphi< 2\pi$) \cite{Voelk:11a}.
\\
\begin{figure}[thb]
\includegraphics[width=.90\columnwidth]{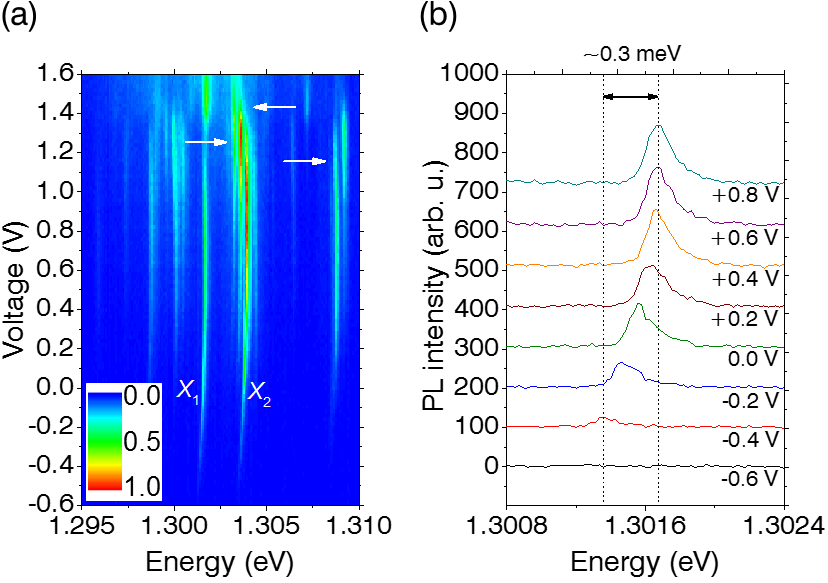}%
\caption{(Color online) (a) False-color plot of bias voltage dependent single QD emission spectra showing a clear QCSE and charging (marked by arrows). 
(b) Selected spectra of the $X_1$ emission for different bias voltages revealing a tuning range of \unit[0.3]{meV} of the QDs in our samples by the QCSE.} \label{FIG2}
\end{figure}

As a first step, we investigate the \emph{static} bias voltage tuneability of single QDs in the epitaxially transferred \emph{p-i-n}-diode. Typical PL spectra of a single QD are plotted in false color representation as a function of the applied bias voltage $(V_B)$ in Fig. \ref{FIG2} (a). At large negative $V_B<\unit[-0.4]{V}$ no PL is detected due to tunnel extraction of carriers from the QD \cite{Muller2013}. As $V_B$ is increased $\unit[-0.4]{V}\leq V_B\leq\unit[-0.1]{V}$ two prominent emission lines labeled $X_1$ ($E_{X_1}=\unit[1301.6]{meV}$)and $X_2$ ($E_{X_2}=\unit[1303.8]{meV}$) can be distinguished. in the range $\unit[-0.4]{V}\leq V_B\leq\unit[-0.1]{V}$. Both emission lines are observed over a relatively large range of $V_B$ and clear signatures of charging events (marked by arrows in Fig. \ref{FIG2} (b)) are detected for $V_B=\unit[+1.0]{V}$. Such behavior is readily expected for our diode structure due to the large injection barriers for electrons and holes which inhibit resonant tunnel injection\cite{Baier:01}. In addition to the occupancy state control, all emission lines exhibit clear spectral shifts arising from the quantum confined Stark effect (QCSE)\cite{Fry:00a}. Fig. \ref{FIG2} (b) shows selected spectra zooming in to the $X_1$ emission line for $\unit[-0.4]{V}\leq V_B\leq\unit[+0.8]{V},~\Delta V_B= \unit[+0.2]{V}$. From these data we extract a total shift $\Delta E_{\rm QCSE}\simeq\unit[0.3]{meV}$, typical for these types of QDs in this diode structure. \footnote{We want to note that both key characteristics, voltage control of the QD's occupancy state and emission energy, have been observed with fully comparable performance on as-fabricated, but not epitaxially transferred devices.}\\

\begin{figure}[thb]
\includegraphics[width=.90\columnwidth]{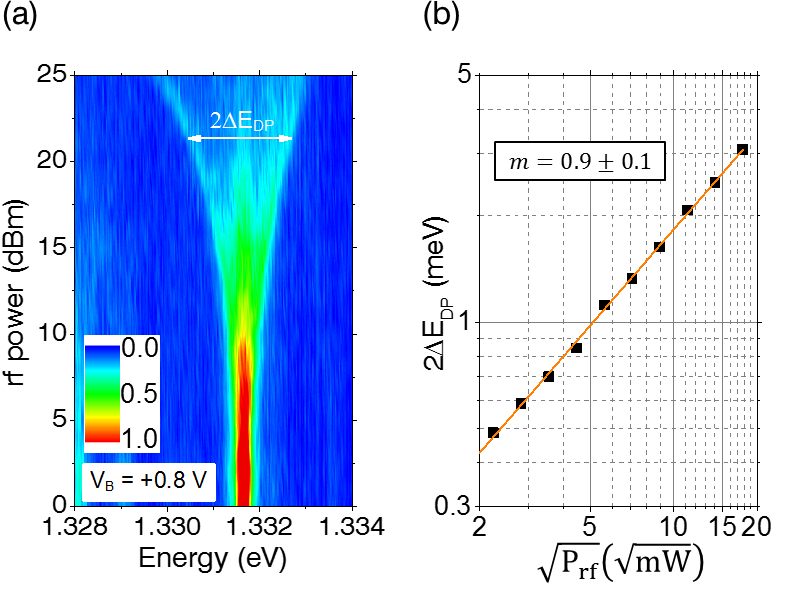}%
\caption{(Color online) (a) False-color plot of a single QD emission line as a function of $P_{rf}$ exhinbiting a pronounced spectral broadening. (b) Extracted spectral shift $\Delta E_{DP}$ as a function of $\sqrt{P_{rf}}\propto A_{SAW}$ revealing a power law $m=0.9\pm0.1$, indicative for DP coupling.} \label{FIG3}
\end{figure} 
In a second step, we assess the \emph{dynamic} acoustic tuning of the QD emission by a SAW. When applying a radio-frequency (rf) signal to the IDT, a SAW is generated on the LiNbO$_3$, which is accompanied by a superposition of an electric and a strain field. As the SAW propagates across the epitaxially transferred photodiode, its electric field component is effectively screened by free carriers in the highly doped layers and the Pd metallization. As a direct consequence the QD response to the SAW is dominated by deformation potential (DP) coupling induced by the dynamic strain field. In Fig. \ref{FIG3} (a) we present \emph{time-integrated, non-stroboscopic} emission spectra of a single QD emission line. \footnote{This particular QD is located underneath a different microaperture on the same photodiode and exhibits a particularly pronounced strain tuning amplitude.} The data are plotted in false color representation as a function of the applied rf power $(P_{rf})$ and a fixed bias voltage of $V_B=\unit[+0.8]{V}$. The emission intensity was averaged over the temporal delay, $\tau_{delay}$, to assess the full tuning bandwidth\cite{Fuhrmann:11}. Clearly, our data demonstrates a pronounced broadening of the emission lines as we increase to $P_{rf} \geq \unit[+4]{dBm}$ \footnote{Due to the pulsed excitation scheme we observe a deviation from the line shape expected for a sinusoidally modulated Lorentzian line since since spectra were averaged over 80 laser pulses, i.e. 80 samples of $\tau_{delay}$. The total tuning bandwidth $(2\Delta E_{DP})$ was determined from the total broadening (low energy to high energy edge) of the emission line \cite{Fuhrmann:11}.}. This broadening and thus spectral modulation induced by DP coupling $(\Delta E_{DP})$ continuously increases with increasing $P_{rf}$. For this particular QD it reaches a maximum of $2\Delta E_{DP}=\unit[3]{meV}$ for the largest SAW amplitudes at $P_{rf} = \unit[+25]{dBm}$. Over this large range of $P_{rf}$ no signatures of a pronounced switching behavior between different emission lines i.e. occupancy states are detected. Such behavior studied in references \cite{Voelk:10b,Schulein2013,Weiss2014b} would be indicative of acousto-electrically driven charge carrier dynamics. Its absence provides evidence for a screening of the SAW-induced electric fields. The nature of the underlying physical mechanism can be identified by studying the modulation bandwidth as a function of the acoustic amplitude $A_{SAW}$. While for DP coupling a linear dependence of $\Delta E_{DP}$ on $A_{SAW}$ is expected, a dynamically driven QCSE should result in a quadratic dependence \cite{Santos2004,Weiss2014a}. To identify such a power law dependence $\Delta E_{DP}\propto A_{SAW}^m$ in our data, we plot the $2\Delta E_{DP}$ extracted from the total width of the emission peak (symbols) over $\sqrt{P_{rf}}\propto A_{SAW}$ in double-logarithmic representation in Fig. \ref{FIG3} (b). This analysis shows a clear linear behavior over the entire range of $P_{rf}$ with no indications of additional contributions. From a best fit (line) we extract an exponent of $m=0.9\pm0.1$, close to the ideal value of $m=1$ expected for DP coupling. Owing the fact that only a single tuning mechanism is at play, we can quantify the local hydrostatic pressure, $p$, dynamically induced by the SAW from $\Delta E_{DP}=\unit[1.5]{meV}$ at $P_{rf} = \unit[+25]{dBm}$. Using the established DP coupling strength in GaAs for [110] and [100] stresses given by the partial derivative of the band gap energy $(E_{gap})$ over the hydrostatic pressure, $\frac{\partial E_{gap}}{\partial p}=\unit[115]{\mu eV/MPa}$\cite{Pollak1968} we obtain a maximum hydrostatic pressure to $p_{max}=\Delta E_{DP}/\frac{\partial E_{gap}}{\partial p}= \unit[13.0\pm 0.6]{MPa}$.\\

\begin{figure}[thb]
\includegraphics[width=.90\linewidth]{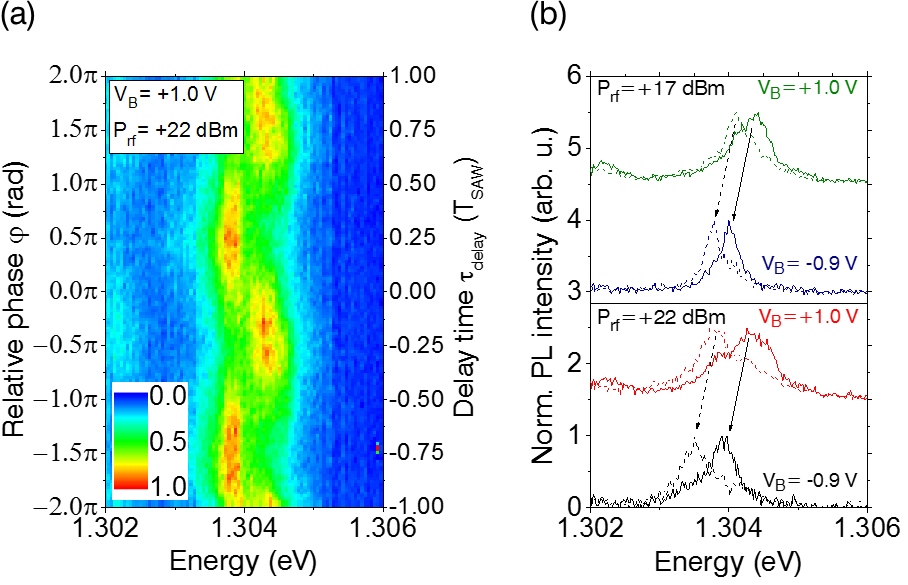}%
\caption{(Color online) (a) Stroboscopic PL spectra of the $X_2$ emission line over two full acoustic cycles resolving the \emph{dynamic nature} of the SAW-driven spectral tuning. (b) Independent control of the $X_2$ emission line by a static electric field and a SAW. The amplitude of the SAW-driven spectral modulation is programmed by $P_{rf}$ (upper and lower panels) and its center energy can be set by $V_B$ as shown by the two sets of stroboscopic spectra in each panel. Respective spectra for maximum compressive (tensile) pressure are plotted as solid (dashed) lines.}\label{FIG4}
\end{figure} 

Finally we address the dynamic nature of the SAW-mediated emission control and its combination with the static electric field tuning. To confirm the time-domain spectral tuning we employed stroboscopic optical excitation. The obtained emission spectra of the $X_2$ emission line for fixed $V_B=\unit[+1.0]{V}$ and $P_{rf}=\unit[+22]{dBm}$ are presented in Fig. \ref{FIG4} (a). The intensity is color coded and plotted as a function of photon energy and $\tau_{delay}$ ($\varphi$) over two full cycles from $-T_{SAW}\leq \tau_{delay} < +T_{SAW}$ ($-2\pi \leq \varphi< 2\pi$). In these data, we resolve a clear spectral oscillation with the fundamental period of the SAW, which exhibits an amplitude of $\Delta E_{DP}=\unit[0.3\pm0.05]{meV}$. These observations are in full agreement with a strain-driven modulation of the QD emission energy and no signatures arising from dynamic piezoelectric effects are resolvable\cite{Weiss2014a}. In particular, the maxima at $\tau_{delay}=-0.75T_{SAW}$ $(\tau_{delay}=+0.25T_{SAW})$ is at lower energies compared to the undisturbed case. This can be attributed to an introduced tensile strain. For $\tau_{delay}=-0.25T_{SAW}$ $(\tau_{delay}=+0.75T_{SAW})$ the energetic shift is towards higher energy, indicative to maximum compressive strain. In the depicted example the peak to peak modulation amplitude is determined to be $2\Delta E_{DP} \approx \unit[0.55]{meV}$.\\
To demonstrate combined static and dynamic tuning employing the QCSE and SAW-driven DP tuning, we compare in Fig. \ref{FIG4}(b) stroboscopic PL spectra recorded of $X_2$ at $V_B=\unit[+1.0]{V}$ and $\unit[-0.9]{V}$ for $P_{rf}=\unit[+17]{dBm}$ (upper panel) and $\unit[+22]{dBm}$ (lower panel). The two stroboscopic spectra were taken at the minimum and maximum tuning of the DP tuning at $\tau_{delay}=-0.75T_{SAW}$ (dashed lines) and $\tau_{delay}=+0.25T_{SAW}$ (solid lines), respectively. Clearly, for both values of $P_{rf}$, tuning of $V_B$ leads to the desired static variation of the center energy of the SAW-driven DP modulation. In addition, the amplitude of the latter is preserved and constant within the resolution of our experiment since for the two chosen values of $\tau_{delay}$ the recorded peak positions exhibit \emph{identical} shifts due to the QCSE as indicated by the dashed and solid arrows. These findings nicely demonstrate \emph{independent} control of single QD emission lines by \emph{static} QCSE and \emph{dynamic} DP tunings. Both parameters are programmed all electrically simply by applying a gate voltage to the photodiode or a rf signal to an IDT to launch a SAW.\\

In conclusion, we realized a LiNbO$_3$-GaAs-hybrid device which enables to deliberately control the optical emission of a single QD by two independently accessible tuning parameters. The unique combination of \emph{dynamic} acousto-mechanical control mediated by a rf SAW and an electro-\emph{statically}, voltage-controlled Stark shift opens directions to add a fast modulation at an arbitrary set transition energy. In our present experiments both ``tuning knobs'' exhibit similar spectral tuning bandwidth. While that of the SAW modulation is mainly dependent on the type of substrate, that of the QCSE-tuning could be dramatically enhanced by introducing AlGaAs barriers\cite{Bennett2010,Trotta2012a} or by replacing the QDs by columnar quantum posts with large QCSE \cite{Krenner:08b}. The demonstrated unique combination of dynamic and static tunings can be employed to implement dynamic quantum gate operations \cite{Blattmann2014} in QD-molecules for which the inter-dot couplings are sensitive to both, electric fields\cite{Krenner:05b} and strain \cite{Zallo2014}. Finally we note, that even large\cite{MichaelisdeVasconcellos2010} electrical contacts allow for rf modulation also of the QCSE. A combination of rf electrical and acoustic offers an alternative route to realize \emph{dynamic} quantum gates employing shaped control pulses\cite{Blattmann2014}.\\ 

We gratefully acknowledge financial support by Deutsche Forschungsgemeinschaft (DFG) within the framework of SFB 631 and the Emmy Noether Program (KR3790/2-1). KM acknowledges financial support from the Alexander von Humboldt Foundation.

%

\end{document}